\documentclass[cits]{PoS}

\usepackage{amssymb,fontenc,times,mathptmx,graphicx}
\usepackage{amsmath,latexsym}
\usepackage{axodraw,mathrsfs}
\usepackage{bbm}
\usepackage{macros}

\title{Analysis of the Schr\"odinger Functional with Chirally Rotated Boundary Conditions} 

\ShortTitle{Analysis of the Schr\"odinger Functional with Chirally Rotated Boundary Conditions}

\author{
\vspace*{-10mm}

\begin{flushright}
HU-EP-08/33\\
SFB/CPP-08-71\\
DESY 08-121\\
LTH 804\\
\end{flushright}
\vspace*{7mm}

J. Gonzalez Lopez$^{b,z}$,
\speaker{K. Jansen}$^z$ and A. Shindler$^l$\\
  \llap{$^b$}Humboldt-Universit\"at zu Berlin, Institut f\"ur Physik\\
  Newtonstrasse 15, 12489 Berlin, Germany\\
  \llap{$^l$}Division of Theoretical Physics, University of Liverpool\\
  Peach Street, Liverpool L69 7ZL, United Kingdom \\
  \llap{$^z$}DESY\\
  Platanenallee 6, 15738 Zeuthen, Germany\\
  E-mail: \email{jenifer.gonzalez.lopez@desy.de}, \email{karl.jansen@desy.de},
  \email{andrea.shindler@liv.ac.uk}
}

\abstract{
The Schr\"odinger functional provides a
valuable tool to perform non-perturbative renormalization on the lattice,
in particular in a mass independent scheme.
We study two different types of chirally rotated Schr\"odinger functional boundary conditions which
have been recently proposed to retain the bulk automatic O(a) improvement
of massless Wilson fermions in finite volume.
We investigate the spectral properties and the quark propagators
which derive from these two proposals in the continuum at tree-level of perturbation theory.
}

\FullConference{The XXVI International Symposium on Lattice Field Theory \\
         July 14 - 19, 2008\\
         Williamsburg, Virginia, USA}

\begin{document}

\section{QCD and tmQCD continuum actions in infinite volume}
\label{sec:tmQCD}

In the continuum, the QCD action with $N_{f}=2$ mass
degenerate quarks in euclidean space and at tree-level of PT takes the form
\begin {equation}
S_{F}[\bar{\psi},\psi]=\int d^{4}x \,
\bar{\psi}(x) \, D\, \psi(x) \quad ; \quad D := \gamma_{\mu}\partial_{\mu} + m.  
\end{equation}
Performing the non-singlet axial rotation on the quark fields,
\begin{equation}
\label{eq:axialrotation}
\psi(x) = e^{i\frac{\alpha}{2}\gamma_{5}\tau^{3}}\, \chi(x)\; , \qquad \bar{\psi}(x) = \bar{\chi}(x)\, e^{i\frac{\alpha}{2}\gamma_{5}\tau^{3}},
\end{equation}
which relates the so called physical, $\{\psi,\bar{\psi}\}$, and twisted bases, $\{\chi,\bar{\chi}\}$,
the form of the tmQCD action is obtained
\begin{equation}
S_{F}^{tm}[\bar{\chi},\chi]=\int d^{4}x \,
\bar{\chi}(x) \, D_{tm} \, \chi(x) \quad ; \quad D_{tm} := \gamma_{\mu}\partial_{\mu} + m_{q} + i\mu_{q}\gamma_{5}\tau^{3} 
\end{equation}
\begin{equation}
m_{q} := m \cos(\alpha) \quad \mu_{q} := m \sin(\alpha). 
\end{equation}
The mass term is now given by the sum of two terms, the untwisted, $m_{q}$, and the twisted, $\mu_{q}$, quark mass.
QCD and tmQCD are equivalent in the continuum, in the sense that they describe the same physics.
This is due to the invariance of the functional integral under the axial rotation given
in Eq.~\eqref{eq:axialrotation}. They share all the symmetries and in particular, $\mathcal{C}\mathcal{P}\mathcal{T}$ is a symmetry.\\
The choice of the rotation angle
$\alpha = \frac{\pi}{2}$ (or equivalently $m_{q}=0, m=\mu_{q}$),
so called maximal twist, is of particular interest since,
when regularizing the theory on the lattice with the standard Wilson term, tmLQCD,
the observables will be automatically O$(a)$-improved~\cite{FR1}.

\section{Motivating chirally rotated Schr\"odinger functional  boundary conditions}
\label{sec:SFbc}

The Schr\"odinger functional (SF) is a finite volume scheme which allows in principle for a non-perturbative (when a lattice regulator is chosen)
and mass-independent renormalization of the theory~\cite{luescher1,karl}.
When appropriate boundary conditions are chosen, it enables
to perform lattice simulations at or close to the chiral point.
Moreover, it allows to study non-perturbatively the
scale dependence of the coupling or renormalization constants
over a wide range of energies,
connecting perturbative and non-perturbative regimes of QCD. 
The SF for pure gauge theory has been introduced in~\cite{luescher1}, and for QCD in~\cite{sint1}.
The theory is defined in a four dimensional Euclidean space where in one of the four directions, which is conventionally chosen to be the
time direction, Dirichlet boundary conditions (b.c.) are imposed.
In the remaining spatial directions periodic boundary conditions (up to a phase) are chosen.\\
The drawback of SF schemes
is the presence of O$(a)$ boundary effects,
even when the lattice theory with e.g. periodic b.c. is O$(a)$-improved.
In the case of Wilson quarks, also bulk O$(a)$ effects will be present
if standard SF boundary conditions are considered,
even in the chiral limit which is expected to be O$(a)$-improved at finite volume.
The desire to retain bulk automatic O$(a)$-improvement with Wilson fermions~\cite{FR1}
has motivated a chiral rotation of the SF b.c.  
Two proposals for chirally rotated SF b.c. are at the moment in the literature,
the twisted SF b.c. of ref.~\cite{sint2} (tSF) and the twisted SF b.c. of ref.~\cite{FR2} ($\gamma_{5}$SF).\\ 
To scrutinize the different proposals, the first step should be to analyze the continuum target theory at tree-level of PT.
In the following
we present our results from the study of the eigenvalue problem and the quark propagator in the presence of the different SF b.c. See also the contributions from S.~Sint and B.~Leder to Lattice 2008. 

\section{Standard and twisted SF boundary conditions}
\label{sec:boundariesSint}

The standard SF boundary conditions~\cite{sint1,luescher2} are given by the equations
\footnote{All the definitions concerning the discrete symmetries can be found in ref.~\cite{andrea}.}
\begin{align}
P_{+}\psi(x)|_{x_{0} = 0}& =0&
P_{-}\psi(x)|_{x_{0} = T}& =0 \quad \textrm{via }\mathcal{T}\\
\bar{\psi}(x)P_{-}|_{x_{0} = 0}& =0 \quad \textrm{via }\mathcal{C}&
\bar{\psi}(x)P_{+}|_{x_{0} = T}& =0 \quad \textrm{via }\mathcal{T}\textrm{ and }\mathcal{C}
\end{align}
where, `via $\mathcal{C}$ or $\mathcal{T}$' means that this b.c. are obtained from the b.c.
for the quark field $\psi(x)$ at $x_{0}=0$ using the transformations $\mathcal{C}$ and/or $\mathcal{T}$
and
\begin{equation}
P_{\pm} = \frac{1}{2}\, \left( \mathbbm{1} \pm \gamma_{0} \right).
\end{equation}
Performing the axial rotation~\eqref{eq:axialrotation} on the quark fields, in the continuum,
the SF b.c. are twisted to the form~\cite{sint2}
\begin{align}
P_{+}(\alpha)\chi(x)|_{x_{0} = 0}& =0&
P_{-}(\alpha)\chi(x)|_{x_{0} = T}& =0\\
\bar{\chi}(x)\gamma_{0}P_{-}(\alpha)|_{x_{0} = 0}& =0&
\bar{\chi}(x)\gamma_{0}P_{+}(\alpha)|_{x_{0} = T}& =0
\end{align}
\begin{equation}
P_{\pm}(\alpha) = \frac{1}{2}\, \left( \mathbbm{1} \pm \gamma_{0}\, e^{i\alpha\gamma_{5}\tau^{3}} \right)
\end{equation}
which for maximal twist setup, $\alpha = \frac{\pi}{2}$, take the form (tSF b.c.)
\begin{align}
Q_{+}\chi(x)|_{x_{0} = 0}& =0&
Q_{-}\chi(x)|_{x_{0} = T}& =0 \quad \textrm{via }\mathcal{T}_{F}^{1,2}\\
\bar{\chi}(x)Q_{+}|_{x_{0} = 0}& =0 \quad \textrm{via }\mathcal{C}&
\bar{\chi}(x)Q_{-}|_{x_{0} = T}& =0 \quad \textrm{via }\mathcal{T}_{F}^{1,2}\textrm{ and }\mathcal{C}
\end{align}
with projector
\begin{equation}
Q_{\pm} := P_{\pm}(\pi/2) = \frac{1}{2}\, \left( \mathbbm{1} \pm i\gamma_{0}\gamma_{5}\tau^{3} \right).
\end{equation}
It is important to notice the correspondence between the relation SF-tSF and QCD-tmQCD.
The discrete symmetries of the SF are the same as those of QCD
while the discrete symmetries of the tSF correspond to the ones of tmQCD at maximal twist.

\subsection{Eigenvalue Spectrum}
\label{sec:eigenvaluesStandardSint}

In this section we summarize the work done in \cite{sint1} for SF b.c. and
we perform the same study for the tSF b.c. proposed in~\cite{sint2}.
In particular, we want to analyze whether the newly proposed SF b.c. retain the gap
in the eigenvalue spectrum.

\subsubsection{Standard SF boundary conditions}
\label{sec:eigenvaluesStandard}

Due to the presence of the SF b.c., $\psi$ and $\bar{\psi}^{\dagger}$ belong to different vector spaces.
In order to be able to write the action as a quadratic form and have a well defined eigenvalue
problem, the complementary components of the quark fields must satisfy Neumann b.c.
\begin{align}
(\partial_{0} - m)P_{-}\psi(x)|_{x_{0} = 0}=0,\quad (\partial_{0} + m)P_{+}\psi(x)|_{x_{0} = T}=0\\
\bar{\psi}(x)P_{+}(\partial_{0} + m)|_{x_{0} = 0}=0,\quad \bar{\psi}(x)P_{-}(\partial_{0} - m)|_{x_{0} = T}=0.
\end{align}
The finite size of the space imply a discrete spectrum of the Dirac operator
$D^{\dagger}D$, which due to the structure of these particular b.c. has a
non-zero lower bound
\be
\lambda_{0}^{2}(m=0)= \left(\frac{\pi}{2\textrm{T}}\right)^{2}.
\ee
It can be seen from this equation how the bound in the spectrum
originates from the b.c. in the time direction
since it is the time extent of the system, $\textrm{T}$, which provides the spectral gap.
The existence of this bound is a crucial point if we are interested
in a mass independent scheme and regularizing the
theory on the lattice, where the numerical inversion of $D^{\dagger}D$ is required
even in the massless limit.

\subsubsection{Twisted SF boundary conditions}
\label{sec:eigenvaluesSint}

The tSF b.c. imply in this case that $u$ and $d$ fields belong to different vector spaces
while $u$ and $\bar{u}^{\dagger}$ ($d$ and $\bar{d}^{\dagger}$) belong to the same space.
Requiring a quadratic form of the action will need the complementary components of the quark fields
to satisfy also Neumann b.c.
\begin{equation}
(\partial_{0} -  \mu_{q})Q_{-}\chi(x)|_{x_{0} = 0}=0,
\quad (\partial_{0} + \mu_{q})Q_{+}\chi(x)|_{x_{0} = T}=0.
\end{equation}
Again, the discrete spectrum of the Dirac operator, $D_{tm}^{\dagger}D_{tm}$, has a non-zero lower bound
\be
\label{eq:eigenvaluessint}
\lambda_{0}^{2}(\mu_{q}=0)=\left(\frac{\pi}{2\textrm{T}}\right)^{2} + m_{q}^{2}.
\ee
It is important to note that the eigenvalue problem with tSF b.c. has an identical structure compared with the SF b.c.;
only the role of the twisted and untwisted quark masses has been switched
(and the twisted projectors have been used).
Therefore, looking at Eq.~\eqref{eq:eigenvaluessint} it can be seen that the only effect of the untwisted mass term is
to lift the eigenvalues, while all mass dependence of the structure of the eigenvalue
is given by the twisted mass term.\\
Only at maximal twist, i.e. at zero untwisted quark mass,
the eigenvalues have exactly the same form as with the standard SF b.c.
(with of course $\mu_q$ instead of $m$),
$\lambda_{n}^{2}(m) = \lambda_{n}^{2}(\mu_{q}) \quad \forall n$.
This is the expected result since the tSF b.c. are obtained from a maximal twist rotation of the SF b.c.

\subsection{Quark propagator}
\label{sec:spectrumStandardSint}

Due to the b.c. on the quark fields,
the quark propagator in the theory with SF b.c. is a solution of the equations
\begin{align}
  &D\left(x \right)\,S^{\textrm{SF}}\left( x,y \right) = \delta^{4}\left( x-y\right) \; , \qquad 0 < x_{0},y_{0} < T\\
  &P_{+}S^{\textrm{SF}}(x,y)|_{x_{0} = 0} =0 \; , \qquad P_{-}S^{\textrm{SF}}(x,y)|_{x_{0} = T} =0
  \label{eq:boundariesSstandard}
\end{align}
while the quark propagator in the theory with tSF b.c. is obtained from
\begin{align}
  &D\left(x \right)\,S^{\textrm{tSF}}\left( x,y \right) = \delta^{4}\left( x-y\right) \; , \qquad 0 < x_{0},y_{0} < T\\
  &Q_{+}S^{\textrm{tSF}}(x,y)|_{x_{0} = 0} =0 \; ,  \qquad Q_{-}S^{\textrm{tSF}}(x,y)|_{x_{0} = T} =0
  \label{eq:boundariesSsint}
\end{align}
with $D(x)$ the corresponding Dirac operator in each case.\\
Additionally, the b.c. on the right side must be satisfied which are given, respectively, by
\begin{equation}
  S^{\textrm{SF}}(x,y)P_{-}|_{y_{0} = 0} =0 \; ,  \qquad S^{\textrm{SF}}(x,y)P_{+}|_{y_{0} = T} =0
\end{equation}
\begin{equation}
  S^{\textrm{tSF}}(x,y)Q_{+}|_{y_{0} = 0} =0 \; ,  \qquad S^{\textrm{tSF}}(x,y)Q_{-}|_{y_{0} = T} =0.
\end{equation}
Unique and non-trivial propagators exist in each case.
The propagator for standard SF b.c., $S^{\textrm{SF}}(x,y)$,
is given in ref.~\cite{luescher3}.
The corresponding propagator $S^{\textrm{tSF}}(x,y)$
is related to the standard one by the non-singlet axial rotation
\eqref{eq:axialrotation} at maximal twist ($\alpha=\pi/2$)
\be
S^{\textrm{tSF}}(x,y) = e^{-i\frac{\pi}{4}\gamma_{5}\tau^{3}}\,
S^{\textrm{SF}}(x,y)\, e^{-i\frac{\pi}{4}\gamma_{5}\tau^{3}}.
\ee

\section{$\gamma_{5}$SF boundary conditions}
\label{sec:boundariesFR}

The $\gamma_{5}$SF b.c. proposed in~\cite{FR2} are also defined for a two-flavour theory and are given by
\begin{align}
\Pi_{+}\phi(x)|_{x_{0} = 0}& =0&
\Pi_{-}\phi(x)|_{x_{0} = T}& =0 \quad \textrm{via }\mathcal{T}\\
\bar{\phi}(x)\Pi_{-}|_{x_{0} = 0}& =0 \quad \textrm{via }\mathcal{C}_{F}^{1,2}&
\bar{\phi}(x)\Pi_{+}|_{x_{0} = T}& =0 \quad \textrm{via }\mathcal{T}\textrm{ and }\mathcal{C}_{F}^{1,2}
\end{align}
\begin{equation}
\Pi_{\pm} = \frac{1}{2}\, \left( \mathbbm{1} \pm \gamma_{5} \tau^{3} \right).
\end{equation}
Although $\mathcal{C}\mathcal{P}\mathcal{T}$ is still a symmetry,
the discrete symmetries separately, $\mathcal{C}_{F}^{1,2}$, $\mathcal{P}_{F}^{1,2}$ and $\mathcal{T}$, with
\begin{displaymath}
  \mathcal{C}_{F}^{1,2}:\left\{ \begin{array}{ll}
      \phi(x)\rightarrow i\tau^{1,2}C^{-1}\, \bar{\phi}(x)^{T}\\
      \bar{\phi}(x)\rightarrow \phi(x)^{T}\, i\tau^{1,2}C
    \end{array} \right. 
\end{displaymath}
are now different from the discrete symmetries of both QCD and tmQCD.
Moreover, in this case there is no transformation in the continuum
which brings SF to $\gamma_{5}$SF.

\subsection{Eigenvalue Spectrum}
\label{sec:eigenvaluesFR}

Due to the boundaries, the $u$ and $d$ fields belong to different vector spaces
and the $u$($d$) and $\bar{d}^{\dagger}$($\bar{u}^{\dagger}$) fields are in the same space.
In this case, differently to what happens with the two previous SF b.c.,
to demand a quadratic form of the action (for each flavour) and a well defined eigenvalue problem
does \emph{not} imply Neumann b.c. for the complemantary components of the quark fields but
\begin{align}
  &\left( m_{q} + i\mu_{q}\gamma_{5}\tau_{3} \right)\Pi_{-}\phi(x)|_{x_{0} = 0} = 0\label{eq:1}\\
  &\left( m_{q} + i\mu_{q}\gamma_{5}\tau_{3} \right)\Pi_{+}\phi(x)|_{x_{0} = T} = 0\label{eq:2}.
\end{align}
For a non-zero value of the quark mass, homogeneous Dirichlet b.c. are satisfied
thus implying that the quark field is zero everywhere, $\phi(x)=\bar{\phi}(x)=0 \: \forall x$.
Therefore, the only possible non trivial solution could be obtained at zero mass.
However, in the chiral limit no additional b.c. for the quark fields occur.
Due to the lack of additional b.c. there is no condition which constraints the possible values of $p_{0}$
and it leads to a gapless spectrum.
We conclude that the eigenvalue problem with $\gamma_5$SF b.c. has either the trivial 
solution or, if both $m_q$ and $\mu_{\rm q}$ vanish, a gapless spectrum.\\
There are two important remarks which have to be made here.
The first is that from Eqs.~\eqref{eq:1}-\eqref{eq:2} we would arrive to the same conclusion
independently of the form of the mass term.
The second is that, this result is a consequence of the fact that there
is no distinction between the normal ($\gamma_{0}$)
and the tangential ($\gamma_{k}$) components of the fields at the boundaries ($x_{0}=0,T$),
with respect to the projector here considered (note that the projector does not contain
$\gamma_{0}$ in this case).

\subsection{Quark propagator}
\label{sec:spectrumFR}

The equations for the quark propagator are in this case
\begin{align}
  &D\left(x \right)\,S\left( x,y \right) = \delta^{4}\left( x-y\right) \qquad 0 < x_{0},y_{0} < T\\
  &\Pi_{+}S(x,y)|_{x_{0} = 0} =0 \qquad \Pi_{-}S(x,y)|_{x_{0} = T} =0. 
  \label{eq:boundariesSfr}
\end{align}
However, the propagator obtained from these equations~\cite{ajk} does not satisfy the corresponding b.c. on the right side,
\be
S(x,y)\Pi_{-}|_{y_{0} = 0} \ne 0 \qquad S(x,y)\Pi_{+}|_{y_{0} = T} \ne 0.
\ee
This means the only solution for the quark propagator, with quark fields obeying $\gamma_{5}$SF b.c.
is the trivial one.\\
On the contrary, it satisfies different b.c. which are obtained from the b.c. on the left using charge conjugation
\begin{equation}
  S(x,y)\Pi_{+}|_{y_{0} = 0} =0 \qquad S(x,y)\Pi_{-}|_{y_{0} = T} =0.
\end{equation}
The corresponding b.c. for the quark fields would be
\be
\label{eq:bdfrc2}
\bar{\phi}(x)\Pi_{+}|_{x_{0} = 0} =0 \qquad \bar{\phi}(x)\Pi_{-}|_{x_{0} = T} =0,
\ee
inducing a theory with b.c. which violate $\mathcal{P}$ and $\mathcal{C}\mathcal{P}\mathcal{T}$.\\
To conclude, continuum QCD at tree-level of P.T.
with the original $\g5$SF b.c. proposed in~\cite{FR2} has a quark propagator which vanishes everywhere.
A non vanishing solution can be found only if we preserve charge conjugation symmetry among the b.c.
Given the fact that the $\gamma_5$SF b.c. violate parity and preserve time reversal,
we end up with a QCD theory with boundaries which violates $\mathcal{C}\mathcal{P}\mathcal{T}$.\\
We remark that for the tSF b.c. of ref.~\cite{sint2} the situation is different.
It is sufficient to consider parity and time-reversal symmetries in the twisted
basis to see that they actually preserve separately $\mathcal{C}$, $\mathcal{P}$ and $\mathcal{T}$.

\section{Conclusions}
\label{sec:conclusion}

In this proceedings contribution we have investigated, at tree-level of perturbation theory
in the continuum, important aspects of three different ways to implement Schr\"odinger 
functional boundary conditions by analyzing the eigenvalue spectrum and the quark propagator.\\
Our conclusion is that the standard~\cite{luescher1,sint1} and twisted~\cite{sint2} SF b.c
are a sound definition of QCD with SF boundaries,
while the lattice formulation 
proposed in~\cite{FR2} has still open questions
which need to be further investigated
before these type of b.c. can be used in practical simulations.

\acknowledgments

We thank R. Frezzotti, G. C. Rossi and S. Sint for valuable discussions.
J.G.L. thanks the SFB-TR9 for financial support.

\end{document}